\documentclass[seceq,supplement]{ptptex}
\usepackage{graphicx}

\title{
Zero temperature phase diagram of finite connectivity spin glasses}

\author{
Florent \textsc{Krz\c{a}ka{\l}a}%
}

\inst{
Dipartimento di Fisica, INFM and SMC, Universit\`a di
Roma ``La Sapienza'', P.~A.~Moro 2, 00185 Roma, Italy}

\abst{The zero temperature phase diagram of spin glasses on finite
connectivity graphs is investigated, with or without magnetic field
and/or ferromagnetic bias, for mean field (using the cavity method)
and Edwards-Anderson (using numerical ground states computations)
models. In the mean field case we show that the phase diagram is
complex and compute the equivalent of the zero temperature de
Almeida-Thouless line. In the $3d$ model however, we found a trivial
phase diagram. This paper presents new analytical results as well as a
rapid review of numerical works on ground states.}

\begin{document}

\maketitle

Since its proposal in 1975\cite{EA}, the Edwards-Anderson (EA) model
of spins glasses (SG)\cite{BOOKsg,BOOK} --- an Ising model with
ferromagnetic {\it and} anti-ferromagnetic interactions --- has been
the subject of many studies and controversies\cite{Zul}. A simple
question such as the qualitative shape of the phase diagram is still
matter of debates\cite{dAT}. Here, we investigate the generic phase
diagram of finite connectivity spin glasses at zero temperature, with
magnetic field and/or ferromagnetic bias. It is motivated by the
recent progresses in both numerical studies of $T=0$ spin glasses
(mainly obtained by borrowing tools from computer science such as
combinatorial optimization\cite{HoudayerMartin}) as well as in
analytical studies of finite connectivity mean field
systems\cite{Cavity} that last years have witnessed, thus allowing for
a direct comparison. In the following, we first consider spin glasses
on random graphs where new analytical results for the phase diagram
and the zero-temperature equivalent of the de Almeida-Thouless (dAT)
line\cite{dAT} are presented. We then consider $3d$ spin glasses,
using numerical computation of ground states, finding in that case a
trivial phase diagram, without any dAT line.
 
\begin{figure}
  {\includegraphics[width=7.1cm]{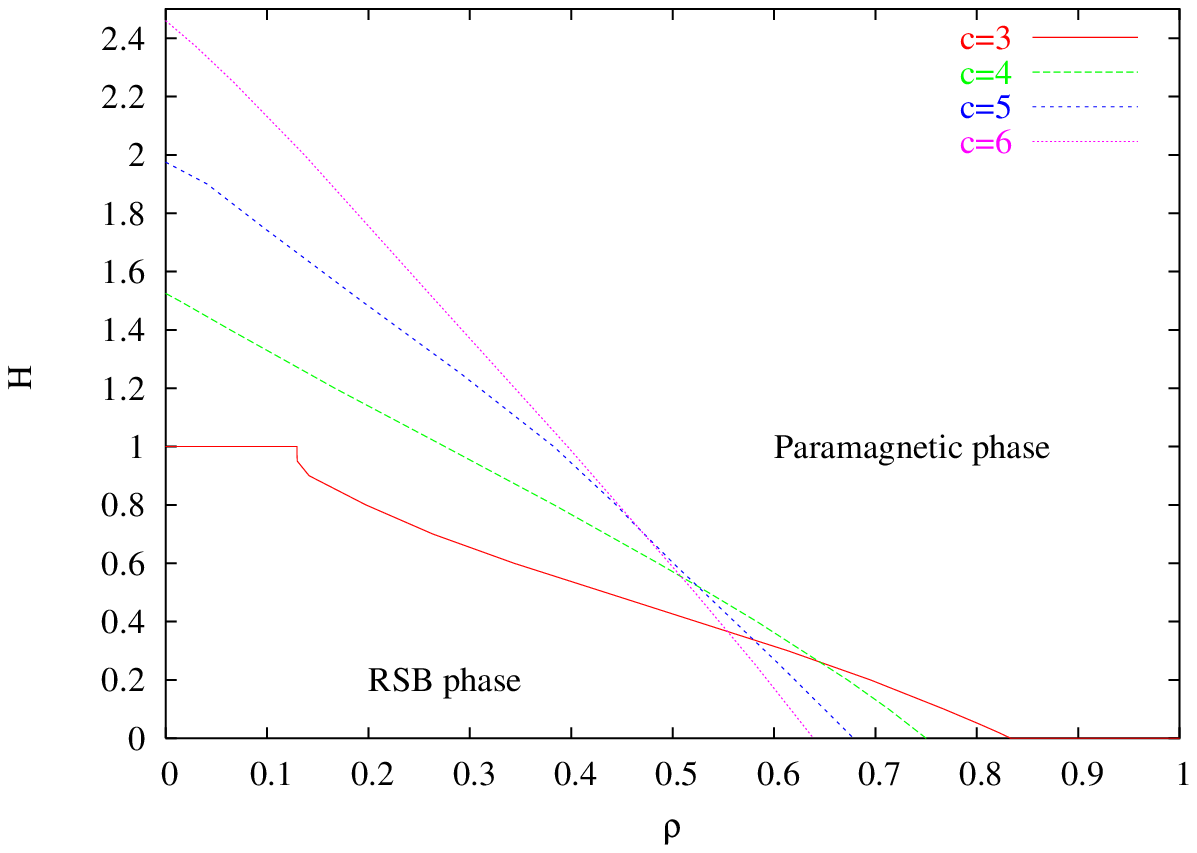}}
  {\includegraphics[width=7.1cm]{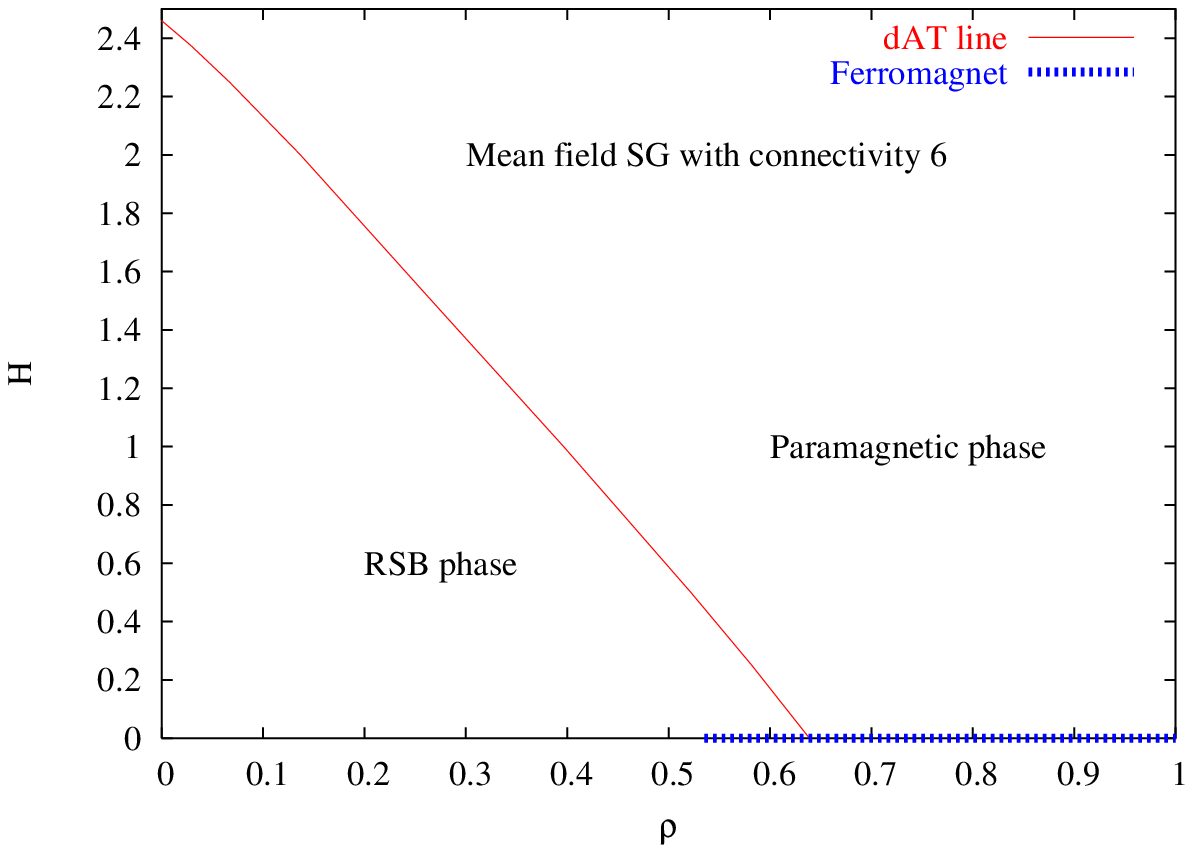}}
  \caption{Zero temperature phase diagram of finite connectivity $c$
mean field spin glasses; $H$ is the applied magnetic field and $\rho$
is the excess concentration in ferromagnetic bonds (i.e. for $\rho=0$
$50$\% of them are ferromagnetic, and for $\rho=1$ we have a pure
ferromagnet). Left: the $T=0$ dAT line separating the paramagnetic
phase and the SG phase for different connectivity in $\pm J$
models. Right: complete phase diagram at connectivity $6$ with the
$T=0$ dAT line and the ferromagnetic phase with spontaneous $m>0$
magnetization: note the presence of a mixed phase with both
ferromagnetic and SG ordering.}
  \label{fig:1}
\end{figure}

{\bf Mean field results}- Most of mean field predictions were derived
within the fully connected Sherrington-Kirkpatrick model using the
replica trick solution of G.~Parisi\cite{BOOK}, characterized by a
so-called Replica Symmetry Breaking (RSB), or many-valleys picture, in
which the energy landscape is divided into many different non ergodic
phases with a complex ultrametric structure\cite{BOOK}. The phase
diagram in presence of a field or with a ferromagnetic bias is quite
rich too: there is a SG phase at low enough field and, when the
concentration of ferromagnetic bonds is enhanced, a 'mixed' phase with
both RSB and spontaneous magnetization appears. However, real spin
glasses {\it do} have a finite connectivity and therefore finite
connectivity mean field models {\it should} be used to study them; the
simplest way of defining them is to consider fixed connectivity random
graphs\cite{Cavity} that we will refer to as Bethe lattices. At $T=0$,
this problem allows many simplifications and computing the phase
diagram using the cavity\cite{Cavity} method turns out to be somehow
easy\footnote{Our $T=0$ computation of the dAT line was made using a
RSB stability analysis equivalent to the {\it bug proliferation}
method\cite{Rivoire} (we send the reader to \citen{Fede} for a similar
computation in the absence of a magnetic field) and independently
confirmed by a numerical 2-replicas cavity analysis {\it a
la} \citen{Pagnani}.}. We present our new results for the phase
diagram of mean field SG with discrete interactions in fig.\ref{fig:1}
where we considered as parameters the applied magnetic field and the
excess concentration of ferromagnetic bonds.
The most important features of this phase diagram are (a) the existence of an
equivalent to the dAT line at $T=0$, that separates the paramagnetic phase at
high field from a SG phase at lower field~\footnote{We computed $H_c$ in many
  cases and found (for $c=6$) $H_c(\rho=0)\approx2.45$ for $\pm J$ couplings,
  $H_c \approx 1.9$ (as \citen{Pagnani}) for Gaussian $J_{ij}$s with uniform
  field, and $H_c \approx 2.4$ for Gaussian $J_{ij}$s with Gaussian random
  field of mean $0$ and standard deviation $H$. For $c=8$, we found (resp.)
  $3.25$, $2.7$ and $3.7$.}  and (b) like in the fully connected model, the
presence of a {\it mixed} phase for high enough concentration of ferromagnetic
bonds, where both a SG ordering with RSB {\it and} a spontaneous magnetization
do set in.  The ferromagnetic ordering was studied here at the RS level, but
RSB corrections can be computed: we have shown recently~\cite{Fede} that in
fact that they even increase the size of the mixed phase\cite{Fede}.

So far we have derived prediction for mean field models; however, and
this is the whole point of the controversy, it is well known that in
the scaling/droplet approach\cite{FH,BrayMoore86} the phase diagram is
trivial (see for instance the Imry-Ma arguments in \citen{FH,MIXED} or
the Migdal-Kadanoff approach of \citen{Migliorini}): there is no dAT
line nor there is a mixed phase and the phase diagram is the same as
for usual ferromagnet. The two theories (mean field vs scaling) are
therefore in conflict and it is still a very debated question to
determine which theory is correct\cite{Zul}. Now that we have a clear
idea of what are their predictions at zero temperature, let us see
what we can say for the $3d$ EA model. To do so, we will now resort to
ground state computations\cite{HoudayerMartin}.

{\bf $3d$ results}- We focus on two questions: (1) Is there a SG phase
when a magnetic field is applied? (2) Is there a mixed phase when the
concentration of ferromagnetic bond is increased?  To detect such a SG
phase in ground state simulations, one has to find a relevant order
parameter. From what we know about spin glasses, a very good candidate
is the following: we say that we have a SG phase if, starting from the
ground state, it is possible to find a system-size excitation (that is
an excitation involving a $O(N)$ spins, $N$ being the size of the
system) of low energy (typically $O(1)$) with a probability constant
(in mean field) or decreasing with the size of the system like a
power-law (in the droplet/scaling approach)\footnote{Note for instance
than in a simple disordered ferromagnet this is going {\it
exponentially} to zero with $N$. For a more detailed discussion, we
refer the reader to \citen{TNT,FH,Thesis}.}. Over the last few years, we
have developed numerical procedures to detect and extract such
excitations, based on the fact that they have a highly non trivial
topology, and we named them {\it sponges}\cite{spo,TNT}. In the
following the probability that a random excitation $O(1)$ energy is a
sponge will be used as the order parameter. Let us now review the
results of our studies.
\begin{figure}
  {\includegraphics[width=7.1cm]{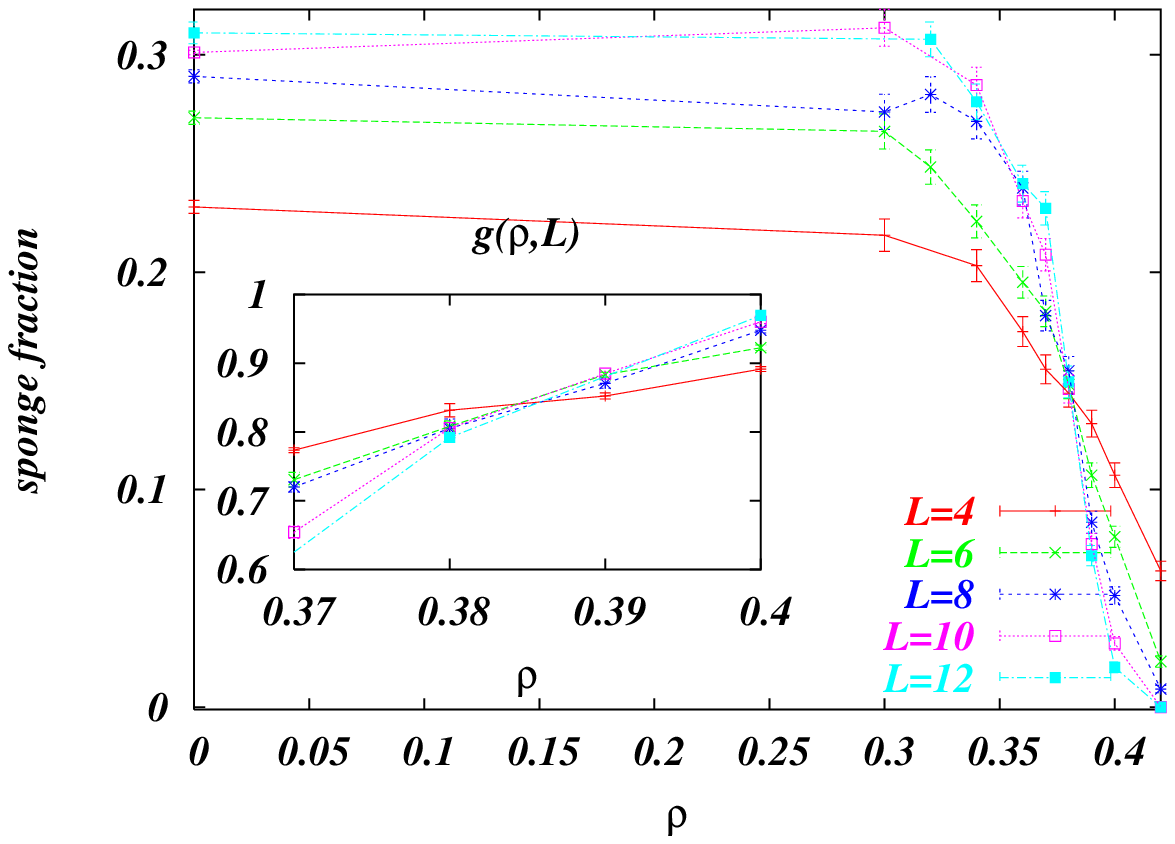}}
  {\includegraphics[width=7.1cm]{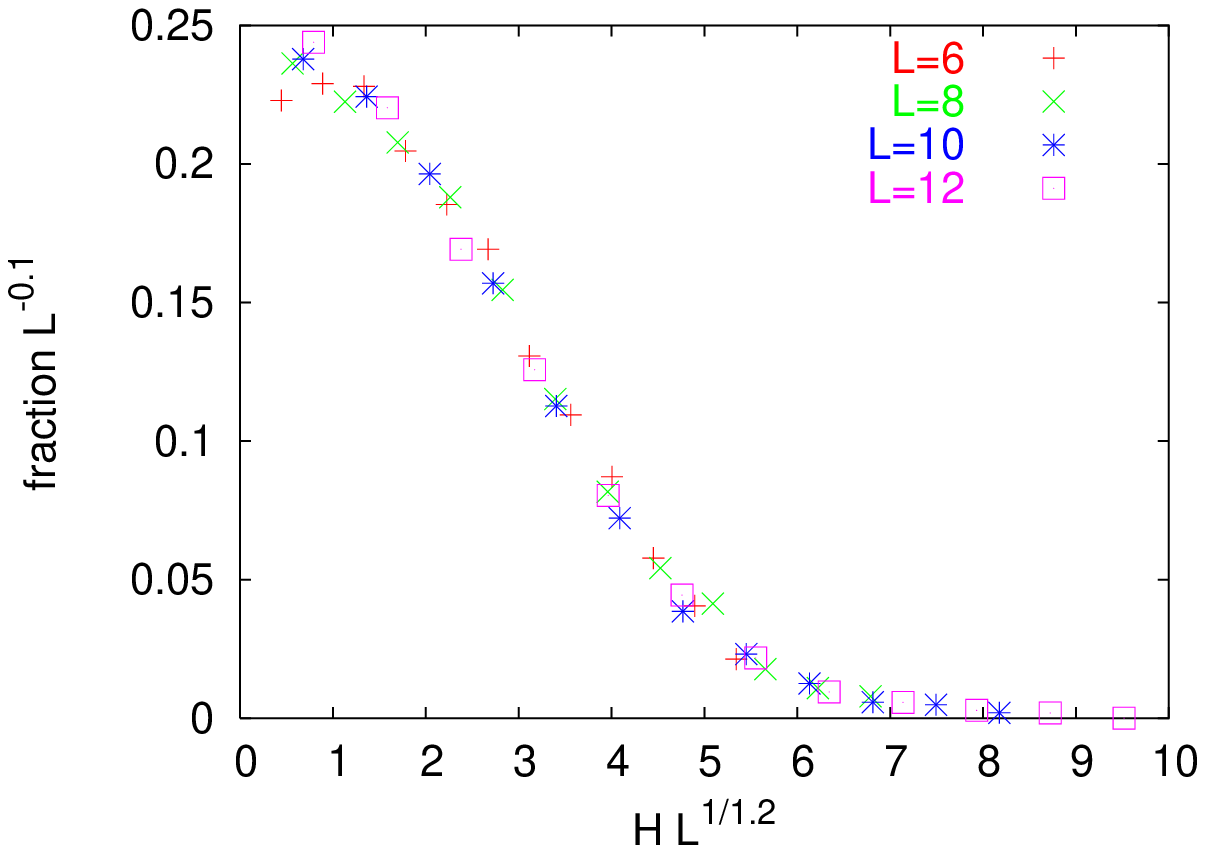}}
  \caption{Results for $3d$ spin glasses. Left: Absence of mixed
  phase; the sponge fraction is vanishing at the same ferromagnetic
  bond ($\approx 0.385$) concentration where a non zero magnetization
  sets in (in inset, the Binder cumulant of the magnetization). Right
  : finite size scaling analysis of the sponge fraction under magnetic
  field, assuming $H_c=0$; data collapse is very good.}
 \label{fig:2}
\end{figure}

First, in the absence of any magnetic field and without any
ferromagnetic bias, we found that the probability of finding such
excitations {\it increases} with the size of the system and seems to
saturates, demonstrating the existence of a SG phase\cite{TNT}. In
fact that would even suggest the existence of a {\it mean field} like
SG phase, but many other mean field ingredients seem to lack: for
instance the fractal dimension of the surface of these excitations is
lower than the space dimension ---the equality $d_s=d$ is fundamental
for mean field predictions--- and we formulated the suggestion that
the whole set of numerical results in $3d$ spin glasses should be
described by a new scenario, that we named TNT, For
Trivial-Non-Trivial\cite{TNT,PY}. Now, how do these results change
when visiting the phase diagram? For the mixed phase, we shown
recently\cite{MIXED} that, when changing the ferromagnetic bond
concentrations, these spongy excitations {\it disappear} at the same
ferromagnetic concentration where the magnetic ordering {\it appears},
thus suggesting that there is {\it no mixed phase} in $3d$ or that it
is at least unobservable (see fig.\ref{fig:2}).

What about the dAT line then? Here finite size effects are more
subtle, making the numerical study more difficult. We argued
in \citen{AT} that the putative critical value is lower than $H_c <
0.65$. To go beyond that, we recently tried\cite{Thesis} a simple
scaling ansatz assuming there is {\it no} dAT line: the (very good)
result is shown in fig.\ref{fig:2} where we used the published data of
the fig.2 in ref \citen{AT}; this strongly suggests an extremely low
value for the critical field, if any (similar conclusions have been
reached by \citen{YK}). In fact, simple droplet arguments\cite{FH}
predict that a rescaling by $L^{-\theta}$ in the $y$ axis and by
$L^{2/(d-2\theta)}$ in the $x$ axis would collapse the curves; taking
the standard value $\theta \approx 0.2$ gives exponents very close to
the one used in fig.\ref{fig:2}. These findings indicate that the
mixed phase or the SG phase in field, if any, happens in an extremely
small range in the phase diagram of the $3d$ EA model. However, we saw
from mean field computation that no such low values should be expected
hence the most direct interpretation is that the $3d$ phase diagram
{\it is} trivial. If similar results were found in $4d$ that would
finally answer negatively the question of a mean field phase diagram
in finite $d$. This is in accord with experimental results\cite{YK} as
well as with the field theory analysis of \citen{Cyrano} where the dAT
line disappears at low dimensions (for $d \leq 6$).

{\bf Conclusion}- Working at $T=0$ allows nice analytical mean field
predictions as well as powerful numerical studies in finite
dimensions. Although the mean field phase diagram is very rich, we
found that the $3d$ one seems trivial. Finally, we want to remark
that, even though many controversies remain\cite{AT,YK}, the TNT
picture, where such a trivial diagram is expected\cite{MIXED},
represents the best resume of the state of art in simulations of $3d$
spin glasses: we see many low energy excitations that however do not
seem to have a mean field structure nor a mean field phase diagram.

 I wish to thank O. C. Martin for the collaboration that led to a
substantial part of the results presented here and acknowledge support
from European Community's Human Potential program under contract
HPRN-CT-2002-00319 (STIPCO).

\end{document}